\documentclass[12pt]{article}
\usepackage{amsmath}
\usepackage{latexsym}
\usepackage{amssymb}
\input{BoxedEPS}

\SetRokickiEPSFSpecial  
\HideDisplacementBoxes

\numberwithin{equation}{section}
\begin{document}

\title{4-Dimensional Einstein Theory Extended by a 3-Dimensional Chern-Simons Term\footnote{QTS3, Cincinnati OH, Sept. 2003}}
\author{R.Jackiw\footnote{Email: jackiw@lns.mit.edu}\\
\\
{\small\itshape Center for Theoretical Physics}
  \\[-1ex]
{\small\itshape Department of Physics}\\[-1ex]
{\small\itshape Massachusetts Institute of
Technology} \\[-1ex]
{\small\itshape  Cambridge, Massachusetts 02139}}

\date{\small MIT-CTP-3428}

\maketitle

\pagestyle{myheadings} \markboth{R.Jackiw}{4-Dimensional Einstein Theory}
\thispagestyle{empty}

\begin{abstract}
When 4-dimensional general relativity is extended by a 3-dimensional gravitational Chern-Simons term an apparent violation of
diffeormorphism invariance is extinguished by the dynamical equations of motion for the modified theory. The physical predictions of this
recently proposed model show little evidence of symmetry breaking, but require the vanishing of the Pontryagin density.
\end{abstract}
\section*{Introduction}
Chern-Simons terms are odd-dimensional entities, interesting to mathematicians \cite{rj1} and physicists.\cite{rj2} For physicists they are
most useful when they are defined on 3-manifolds, and over the last twenty years they have been widely used to model various physical
processes in 3-dimensional space-time, that is, phenomena confined to motion on a plane, like the Hall effect, or  gravitational motion in the
presence of cosmic strings.

However, these 3-dimensional structures can also be inserted into physical theories in 4-dimensional space-time,
and because of the dimensional mismatch this gives rise to kinematical and dynamical violation of Lorentz symmetry, CPT symmetry, etc.

As you have heard earlier in this meeting from Kostelecky, the subject of Lorentz and CPT symmetry violation is interesting these days,
mainly due to his initiating and stimulating work by theorists and experimentalists. The former build plausible extensions of  standard
theories, with small symmetry-breaking terms; the latter perform more and  more precise experiments limiting
the magnitude of such possible terms. Thus far, no evidence for symmetry breaking has been found; indeed,
conventional symmetries are confirmed, with ever-decreasing uncertainty. 

Today, I shall describe a Chern-Simons modification of 4-dimensional gravity theory -- Einstein's general relativity -- and the associated decrease in symmetry.

Actually, more than a decade ago, George Field, Sean Carroll and I investigated a Chern-Simons modification of Maxwell's electromagnetic theory.
So in order to set the stage for the gravitational extension, I shall first review the Maxwell story. \cite{rj3}

\section{Chern-Simons Modification of Maxwell Theory}

The  Chern-Simons term for an Abelian gauge theory on an Euclidean 3-space reads
\begin{equation}
CS (A) \equiv \frac{1}{4} \varepsilon^{ijk} F_{ij} A_k = \frac{1}{2} {\bf A} \cdot {\bf B}.
\label{1.1}
\end{equation}
The first expression is in tensor notation; the second in vector notation, with $\bf B$ being the magnetic field,
${\bf B} = {\bf \nabla} \times {\bf A}$. All indices are spatial $[i,j,k: x, y, z]$.
A related 4-dimensional formula in Minkowski space-time defines the topological Chern-Simons current
\begin{equation}
K^\mu = {^\ast F^{\mu \nu}} A_\nu,
\label{1.2}
\end{equation}
where ${^\ast F^{\mu \nu}}$ is the dual electromagnetic tensor.
\begin{equation}
{^\ast F^{\mu \nu}} = \frac{1}{2} \varepsilon^{\mu \nu \alpha \beta} F_{\alpha \beta}
\label{1.3}
\end{equation}
It is seen that the Chern-Simons term (\ref{1.1}) is proportional to the time $t (\mu=0)$ component of the Chern-Simons current, (\ref{1.2})
with the time dependence suppressed. Also the divergence of the topological current is the topological Pontryagin density.
\begin{equation}
\partial_\mu K^\mu = \partial_\mu ({^\ast F^{\mu \nu}} A_\nu) = \frac{1}{2} {^\ast F^{\mu \nu}} F_{\mu \nu}
\label{1.4}
\end{equation}

In Chern-Simons modified electromagnetism the Chern-Simons term (\ref{1.1}) (with field arguments extended to include $t$)
is added to the usual Maxwell Lagrangian.
\begin{equation}
I = \int d^4  x \left(- \frac{1}{4} \ F^{\mu \nu} \,F_{\mu \nu} + \frac{\mu}{2} \ \bf A \cdot \bf B \right)
\label{1.5}
\end{equation}
Here $\mu$, with dimension of mass, measures the strength of the extension. Formula (\ref{1.5}) may be alternatively presented in covariant
notation, with the help of an external, constant embedding 4-vector $v_\mu$.
\begin{eqnarray}
I= \int d^4  x \left(-\frac{1}{4} \ F^{\mu \nu} F_{\mu \nu} + \frac{1}{2} \  v_\mu {^*F}^{\mu \nu} A_\nu \right) \nonumber\\ [8pt]
v_\mu = (\mu, 0) \qquad \qquad \qquad \qquad \qquad
\label{1.6}
\end{eqnarray}
In spite of the presence of the vector potential, the action is gauge invariant: Under a gauge transformation it changes by a surface term,
since $\partial_\mu {^\ast F^{\mu \nu}} =0$. This can be made explicit by recognizing that in (\ref{1.6}) there occurs the Chern-Simons current
$ K^\mu$ (\ref{1.2}). Therefore, with the help of (\ref{1.4}) and an integration by parts the action acquires a gauge invariant form.
\begin{eqnarray}
I= \int d^4  x \left(-\frac{1}{4} \ F^{\mu \nu} F_{\mu \nu} + \frac{1}{4} \theta \ {^\ast \negthickspace F^{\mu \nu}} F_{\mu \nu} \right)
\nonumber\\ [8pt]
\partial_\mu \theta \equiv v_\mu \qquad \qquad \qquad \qquad \qquad
\label{1.7}
\end{eqnarray}
The external quantity is now $\theta$, which is taken as $\theta=\mu t$ so that (\ref{1.5}) and (\ref{1.6}) are reproduced.

Since the explicitly covariant formulations (\ref{1.6}), (\ref{1.7}) involve external, fixed quantities [a fixed constant embedding
vector $v_\mu$ in (\ref{1.6}); a fixed function $\theta$, linear in time, in (\ref{1.7})], we expect that Lorentz invariance is lost.
Also, since ${\bf A \cdot B}$, and ${^\ast F^{\mu \nu}} F_{\mu \nu}$ are axial quantities, parity is lost; but $C$ and $T$
are preserved, so CPT is also lost. To confirm these statements, we now look to the solutions of the modified equations of motion.

In the electromagnetic equations of motion, which follow from the Chern-Simons extended action, only Amp\`{e}re's law is modified.
\begin{equation}
-\frac{\partial {\bf E}}{\partial t}  + \nabla {\bf \times} \ {\bf B} = {\bf J} + \mu {\bf B}
\label{1.8}
\end{equation}
All other Maxwell equations continue to hold. Also the consistency condition on  (\ref{1.8}) remains as in Maxwell theory: 
the charge density $\rho = {\bf \nabla} \cdot {\bf E}$ and the current $\bf J$ must satisfy their continuity equation, as is seen by taking
the divergence of (\ref{1.8}) and using ${\bf \nabla \cdot B}=0$.

The modification that we have constructed is particularly felicitous for the following reasons.
\begin{description}
\item{(i)}
Gauge invariance is maintained, so the photon continues to possess just two independent polarizations.
\item{(ii)}
Eg.(\ref{1.8}) is not a radical departure; it has played previous roles in physical theory: in plasma physics one frequently
replaces the source current $\bf J$ with a magnetic field $\bf B$. Of course, we are not working with a collective/phenomenological theory, like
plasma physics, rather we are examining the  feasibility of (\ref{1.8}) for fundamental physics.
\end{description}

To assess the actual physical content of the Chern-Simons extended electromagnetism, and its associated symmetry breaking,
we have examined some solutions. We found that in the source-free $({\bf J} = {\bf 0})$ case, plane waves continue to solve the extended equations.
The photon posseses two independent polarizations, (as anticipated from gauge invariance) however they travel at velocities which
differ from the velocity of light (thus Lorentz boost invariance is lost---as anticipated) and also the two polarizations
travel with velocities that differ from each other (thus parity invariance is lost--as anticipated).

The fact that the two photon helicities travel (in vacuum) with different velocities makes empty space behave as a birefringent medium.
Consequently linearly polarized light, passing through this birefringent environment, undergoes a Faraday-like rotation, which can be looked
for in observations of light from distant galaxies. Much data exists on this phenomenon, and the conclusion is unavoidable: there is no such
effect in Nature;
$\mu=0$ is required. This was asserted in our initial investigations \cite{rj3}, \cite{rj4}, and the many other analyses carried out in the
intervening years support that conclusion (see e.g. \cite{rj5}).

\section{Chern-Simons Modification of Einstein Theory}
{\bf A. Gravitational Chern-Simons term in 3-space}.\\ 
The 3-dimensional, gravitational Chern-Simons term can be presented in terms
of the 3-dimensional Christoffel connection $^3\Gamma_{i q}^p$, \cite{rj2}
\begin{equation}
CS(\Gamma) =   \varepsilon^{ijk} \ (\frac{1}{2}
\ ^3\Gamma^p_{i q} \, \partial_j  {^3 \Gamma^q_{k p}} + \frac{1}{3}  {^3\Gamma}^p_{iq} \, {^3\Gamma}^q_{j r} \, ^3\Gamma^r_{kp}),
\label{2.1}
\end{equation}
but it is understood that the Christoffel connection takes the usual expression in terms of the metric tensor, which is the fundamental
variable. Variation with respect to the metric tensor of the intergrated Chern-Simons term results in the 3-dimensional ``Cotton tensor";
which involves a covariant curl of the 3-dimensional Ricci tensor $^3R^i_j$.
\begin{equation}
\frac{\delta}{\delta g_{ij}} \int d^3 x CS(\Gamma) =-\sqrt{g} \ {^3C^{ij}} = \frac{1}{2}
\varepsilon^{imn} \ {^3D_m} {^3R^j_n} + i\leftrightarrow j
\label{2.2}  
\end{equation}
${^3C^{ij}}$ is symmetric, traceless and covariantly conserved. It vanishes if and only if the 3-dimensional metric tensor is conformally flat.
A related formula gives the 4-dimensional Chern-Simons current $K^\mu$,
\begin{equation}
K^\mu = 2 \varepsilon^{\mu}  {^{\alpha \beta \gamma}} \left [\frac{1}{2} ~ \Gamma^\sigma_{\alpha \tau} \, \partial_ \beta
~ \Gamma^\tau_{\gamma \sigma} + \frac{1}{3} ~  \Gamma^\sigma_{\alpha \tau} \,  \Gamma^\tau_{\beta \eta} \Gamma^\eta_{\gamma \sigma} \right],
\label{2.3}
\end{equation}
whose divergence is the topological Pontryagin density.
\begin{equation}
\partial_\mu K^\mu = \frac{1}{2} {^*R}^\sigma_{~\tau} \ {^{\mu \nu}} \ R^\tau {_{\sigma \mu \nu}} \equiv \frac{1}{2} {^*RR}
\label{2.4}
\end{equation}
Here $R^\tau_{~\sigma \mu \nu}$ is the Riemann curvature tensor and $^\ast R^\sigma_{~ \tau} {^{\mu \nu}}$ is its dual.
\begin{equation}
^*R^{\sigma \ \mu \nu}_{\, \,\tau} = \frac{1}{2} ~ \varepsilon^{\mu \nu \alpha \beta} R^\sigma_{~{\tau \alpha \beta}}
\label{2.5}
\end{equation}
[Notation: $(i, j,...)$ are 3-dimensional, spatial indices, and 3-dimensional geometric entities are decorated with the superscript ``3". Undecorated
geometric entities are 4-dimensional, and Greek indices label the 4 space-time coordinates.]
Note that unlike in the vector case, the Chern-Simons term (\ref{2.1}) is not the time component $K^0$, because the former contains
3-dimensional Christoffel entities, while 4-dimensional ones are present in $K^0$. This variety allows various
extensions general relativity.\\ \vspace{-6pt}

{\noindent {\bf B. Gravitational Chern-Simons term in 4-space.}}

In analogy with the electromagnetic formulation (\ref{1.7}), we choose to extend Einstein theory by adopting the
action \cite{rj6}
\begin{eqnarray}
I&=& \frac{1}{16 \pi G} \int d^4 x \left( \sqrt{-g} R + \frac{1}{4} \theta {^*RR} \right) \nonumber \\ [8pt]
&=& \frac{1}{16 \pi G} \int d^4 x \left(\sqrt{-g} R - \frac{1}{2} v_\mu K^\mu \right), \qquad v_\mu \equiv \partial_\mu \theta.
\label{2.6}
\end{eqnarray}
The first contribution is the usual Einstein-Hilbert term involving the Ricci scalar $R$. The modification involves an external
quantity: $\theta$ in the first equality; $\partial_\mu \theta \equiv v_\mu$ in the second equality, which follows from the first by (\ref{2.4})
and an integration by parts. Eventually we shall take the embedding vector $v_\mu$ to possess only a time component, and $\theta$ to depend
solely on time. So then our modification (\ref{2.6}) involves the time component of 4-dimensional Chern-Simons current (\ref{2.3}) 
[rather than the 3-dimensional Chern-Simons term (\ref{2.1})].

The equation of motion that emerges when (\ref{2.6}) is varied with respect to $g_{\mu \nu}$ is
\begin{equation}
G^{\mu \nu} + C^{\mu \nu} = -8 \pi G T^{\mu \nu}.
\label{2.7}
\end{equation}
Here $G^{\mu \nu}$ is the covariantly conserved (Bianchi identity) Einstein tensor, $G^{\mu \nu} = R^{\mu \nu} -\frac{1}{2} g^{\mu \nu} R$.
We have inserted a source with strength $G$ (Newtons constant) consisting of the matter energy-momentum tensor
$T^{\mu \nu}$, which also is convariantly conserved, since we assume matter to be conventionally, covariantly coupled to gravity.
$C^{\mu \nu}$ is the term with which we are extending the Einstein theory.
\begin{equation}
\sqrt{-g} \ C^{\mu \nu} = \frac{\delta}{\delta g_{\mu \nu}} \ \frac{1}{4} \int d^4 x \ \theta
{^\ast RR} = -\frac{1}{2} \left( v_\sigma \varepsilon^{\sigma \mu \alpha \beta} D_\alpha
R^\nu_\beta +  v_{\sigma \tau} {^\ast R^{\tau \mu \sigma \nu}} + \mu \leftrightarrow \nu \right)
\label{2.8}
\end{equation}
$C^{\mu \nu}$ is manifestly symmetric; it is traceless because $^*RR$ is conformably invariant.
$C^{\mu \nu}$'s first term (involving the curl of $R^\nu_\beta$) is similar to the 
3-dimensional $\sqrt{g} {^3C^{i j}}$ (\ref{2.2}). Even the second term can be viewed as a generalization
from 3-dimensions: it involves only the Weyl tensor part of Riemann tensor, which vanishes in 3 dimensions.
[Cotton defined his tensor in arbitrary dimensions $d$, and his definition is equivalent to ours in $d=3$, where it is also given by the variation of
the 3-d gravitational Chern-Simons term, as is (\ref{2.2}). \cite{rj7} However for $d\ne 3$, Cotton's tensor does not appear to have a variational
definition. Our $d=4$ Cotton-like tensor in (\ref{2.8}) does possess a variational definition, at the the expense of introducing non-geometrical
entities like
$\theta \ \mbox{and}\ v_\mu$.]

Finally we must examine $D_\mu C^{\mu \nu}$, whose vanishing is a consistency requirement on (\ref{2.7}).
However, an explicit evaluation (which involves only geometric identities) shows that, unlike in 3 dimensions, $C^{\mu \nu}$ is not covariantly
conserved. Rather
\begin{equation}
D_\mu C^{\mu \nu} = \frac{1}{8 \sqrt{-g}} \ v^\nu {^*RR}.
\label{2.9}
\end{equation}
Thus the vanishing of $^\ast RR$ is a consistency condition of the new dynamics: every solution to (\ref{2.7})
 will necessarily lead to vanishing Pontryagin density.
\begin{equation}
G^{\mu \nu} + C^{\mu \nu} = -8 \pi G T^{\mu \nu} \Rightarrow {^\ast RR} = 0
\label{2.10}
\end{equation}

We may derive and understand the expression for the covariant divergence of $C^{\mu \nu}$ by examining
the response of our addition to changes in the coordinates. With the infinitesimal transformation
\begin{equation}
\delta x^\mu = -f^\mu (x)
\label{2.11}
\end{equation}
we have
\begin{equation}
\delta g_{\mu \nu} = D_\mu f_\nu + D_\nu f_\mu.
\label{2.12}
\end{equation}
The Hilbert Einstein action is of course invariant. To assess the variance properties of our modification,
we can proceed in two ways. First observe that ${^\ast RR}$ is scalar density, so it transforms as
$\delta {(^\ast RR)} = \partial_\mu (f^\mu {^\ast RR})$. $\theta$ is an external quantity, therefore we do not transform
it.
\begin{subequations}
\begin{eqnarray}
\delta I_{CS} = \frac{1}{4} \int d^4 x \ \theta \delta (^\ast RR)&=& \frac{1}{4} \int d^4 x
\theta \partial_\mu (f^\mu {^\ast RR}) \nonumber\\ [8pt] &=&-\frac{1}{4} \int d^4 x
v_\mu f^\mu {^\ast RR}
\label{2.13a}
\end{eqnarray}
Alternatively, we may vary $I_{CS}$, by varying $g_{\mu \nu}$ according to (\ref{2.10}) , and using the definition
(\ref{2.8}) for $C^{\mu \nu}$.
\begin{eqnarray}
\delta I_{CS}&=& \int d^4 y\  \frac{\delta}{\delta g_{\mu \nu} (y)} \ \left(\frac{1}{4} \int d^4 x \ \theta \ {^\ast RR} \right)
2 D_\mu f_\nu (y) \nonumber\\ [8pt]
&=& 2 \int d^4 x \ \sqrt{-g} \ C^{\mu \nu} D_\mu f_\mu = -2 \int d^4 x \ \sqrt{-g}
\ (D_\mu C^{\mu \nu} ) f_\nu 
\end{eqnarray}
\end{subequations}
Equating the two expressions for $\delta I_{CS}$ establishes (\ref{2.9}), and also demonstrates that $^\ast RR$ is a measure of
the failure of diffeomorphism invariance. But $^\ast RR$ vanishes as a consequence of the equation of motion, so
is some sense diffeomorphism invariance is dynamically reinstated.

For another perspective, consider a variant of our model, where $\theta$ in (\ref{2.6}) is  a dynamical
variable, not an externally fixed quantity. If we postulate that under diffeomorphisms (\ref{2.11})
$\theta$ transforms as a scalar,
\begin{equation}
\delta \theta = f^\mu \partial_\mu \theta = f^\mu v_\mu,
\label{2.14}
\end{equation}
 then (2.12a) acquires the additional contribution
\begin{equation}
\frac{1}{4} \int d^4 x \, \delta \theta ~({^\ast RR}) = \frac{1}{4} \int d^4 x \ v_\mu \ f^\mu
\ {^\ast \negthickspace RR},
\label{2.15}
\end{equation}
which cancels (\ref{2.13a}), showing that the Chern-Simons modification with dynamical $\theta$ is
diffeomorphism invariant. Now let us look at the equations of motion in this variant of modified
gravity: varying $g_{\mu \nu}$ still produces (\ref{2.7}); varying $\theta$, which now acts as a Lagrange
multiplier, forces $^\ast RR$ to vanish, but that requirement is already implied by (\ref{2.7}), (\ref{2.10}).
Thus the equations of the fully dynamical, and diffeomorphism invariant theory coincide with the equations
of the non-invariant theory, where $\theta$ is a fixed, external quantity.

Formula (\ref{2.13a}) shows that when $v_\mu$ is chosen to have only a time component, $v_\mu
=(\frac{1}{\mu}, \bf 0)$; equivalently $\theta = t/ \mu$, then $I_{CS}$ is invariant under all space-time
reparametrizations of the spatial coordinates, and also of shifts in time: $f^0 = \mbox{constant},
f^i \mbox{arbitrary}$. Henceforth we make this choice for $v_\mu\ \mbox{and} \ \theta$.\\ 

{\noindent \bf C. Physical effects of the Chern-Simons term in 4-d gravity.}\\
We examine some physical processes in the Chern-Simon modified gravity theory.
\begin{description}
\item{(i)} 
It is important that the  Schwarzschild solution continues to hold; thus our theory  passes the 
``classic" test of general relativity. The result is established in two steps.
First we posit a stationary form for the metric tensor
\begin{equation}
g_{\mu \nu} = \left(
\begin{array}{cc}
N& 0\\
0& g_{ij}
\end{array} \right),
\label{2.16}
\end{equation}
with time-independent entries. It follows that $C^{00}\ \mbox{and}\ C^{n0}= C^{0n}$ vanish. Also one finds
that the spatial components reproduce the 3-dimensional Cotton tensor.
\begin{equation}
\sqrt{-g} \ C^{ij}  =  \sqrt{g}^3 \ C^{ij}
\end{equation}
Next, we make the spherically symmetric {\it Ansatz}, and find that $C^{ij}$ vanishes. Evidently also
$^\ast RR$ must vanish on the Schwarzschild geometry, because the modified equations are satisfied.
Since the Kerr geometry, carries non vanishing $^\ast RR$, it will not be a solution to the extended
equations. It remains an interesting, open question which deformation of the Kerr geometry satisfies
the Chern-Simons modified equations. 
\vspace{8pt}

\item{(ii)} 
Next we perform a linear analysis by expanding the metric tensor around a flat background
$g_{\mu \nu} = \eta_{\mu \nu} + h_{\mu \nu}$. The purpose of the linear analysis is to determine the propagating degrees of freedom, to study
the nature of small disturbances (gravity waves) and to illuminate the construction of an energy-momentum (pseudo)
tensor, which is symmetric and divergence-free.

Keeping only the linear portions of the Einstein tensor and $C_{\mu \nu}$, we
verify that both $G^{linear}_{\mu \nu} \ \mbox{and}\ C^{linear}_{\mu \nu}$ are divergence-free.
\begin{equation}
\partial^\mu G^{linear}_{\mu \nu} = 0 = \partial^\mu C^{linear}_{\mu \nu}
\label{2.17}
\end{equation}
This is seen from the explicit formulas. It also follows from the observation that the exact equation
$D^\mu G_{\mu \nu} =0$ implies the above result for $G^{linear}_{\mu \nu}$; moreover, from (\ref{2.9})
we see that $D^\mu C_{\mu \nu}$ is of quadratic order, hence the above result for $C^{linear}_{\mu \nu}$
holds also.
It is further seen that the linear portions are invariant under the ``gauge" transformation
\begin{equation}
h_{\mu \nu} \to h_{\mu \nu} + \partial_\mu \lambda_\nu + \partial_\nu \lambda_\mu
\label{2.18}
\end{equation}
\end{description}

In the Einstein theory, one decomposes $h_{\mu \nu}$ into temporal parts, and purely spatial parts $h^{ij}$.
The latter is further decomposed into its trace, its longitudinal part, and its traceless transverse part,
denoted by $h^{ij}_{TT}$. One then finds from the linear equations that, with the exception
of $h^{ij}_{TT}$, all other components of $h_{\mu \nu}$ are either non-propagating or can be eliminated
by the gauge transformation (\ref{2.18}). Only $h^{ij}_{TT}$ survives and it is governed by a 
d'Alembertian. Since in 4 dimensions a symmetric, transverse and traceless $3\times 3$ matrix possesses
two independent components, one concludes that in Einstein's theory small gravitational disturbances
are waves, with two polarizations, each moving with the velocity of light (governed by the d'Alembertian).

None of this changes where $C^{linear}_{\mu \nu}$ is included. Again only $h^{ij}_{TT}$ propagates,
governed by the d'Alembertian. Explicitly the modified equation for $h^{ij}_{TT}$ reads
\begin{eqnarray}
(\delta^{im} \delta^{jn} + \frac{1}{2\mu} \varepsilon^{ipm} \ \delta^{nj} \ \partial_p + \frac{1}{2\mu}
\varepsilon^{jpm} \ \delta^{ni} \ \partial_p) \ \Box \ h^{mn}_{TT} \nonumber\\
=-16 \pi \ G \ T^{ij}_{TT}. \qquad \qquad \qquad \qquad \qquad \qquad
\end{eqnarray}
$T^{ij}_{TT}$ is the transverse traceless part of the stress tensor. The new terms are the $(\mu^{-1})$ contributions; they
involve only spatial derivatives. One may consider that the left side of (2.20) involves an operator
acting on $\Box \ h^{mn}_{TT}$.
\begin{subequations}
\begin{equation}
\mathcal{O}^{ij}_{~ \ mn} \Box \ h^{mn}_{TT} = -16 \pi G \ T^{ij}_{TT}
\end{equation}
Acting on this equation with the inverse operator $\mathcal{P} = \mathcal{O}^{-1}$ shows that the effect
of the entire extension is to modify the source
\begin{eqnarray}
\Box \ h^{mn}_{TT}&=&-16 \pi G \ \mathcal{P}^{mn}_{~\ \ \ ij} \ T^{ij}_{TT} \nonumber\\
&\equiv&-16 \pi G \ \tilde{T}^{ij}_{TT}                                                                  
\end{eqnarray}
\end{subequations}

Thus we see that in sharp contrast to the electromagnetic case, the Chern-Simons modification of gravity
does not change the velocity of gravity waves and there is no Faraday rotation. It is also noteworthy
that the reduction to 2 degrees of freedom (2 polarizations) takes place also in the extended theory.
Such a reduction of degrees of freedom is considered to be a consequence of gauge invariance, here diffeomorphism
invariance, which evidently continues to hold on our modified theory.

There does exist a physical manifestation of the extension. Although the velocities of the two polarizations
are the same, their intensities differ, due to the modification of the source $(T^{ij}_{T T} \to \tilde{T}^{ij}_{TT})$.
One finds for  a weak modification (large $\mu$) that the ratio  of the intensity of waves with
negative helicity to those with positive helicity is
\begin{equation}
\frac{-}{+} = \left(1+ \frac{4\omega}{\mu}\right),
\end{equation}
where $\omega$ is the frequency. This puts into evidence the parity violation of the modification.

Finally we turn to the topic of the energy-momentum (pseudo) tensor. A straight forward approach
to this problem in the Einstein theory is to rewrite the equation of motion by decomposing the Einstein
tensor $G_{\mu \nu}$ into its linear and non-linear parts, and moving the non-linear terms to the ``right"
side, summing it with the matter energy-momentum tensor.
\begin{equation}
G^{linear}_{\mu \nu} = -8\pi \ G \left(T_{\mu \nu} + \frac{1}{8\pi  G} \ G^{non-linear}_{\mu
\nu}\right)
\end{equation}
Clearly $G^{linear}_{\mu \nu}$ is symmetric and conserved, therefore, so must be the right side, which is now renamed
as total (gravity + matter) energy-momentum (pseudo) tensor.
\begin{equation}
\tau_{\mu \nu} = T_{\mu \nu} + \frac{1}{8\pi  G} \ G^{non-linear}_{\mu \nu}
\end{equation}

Exactly the same procedure works in the extended theory. We present the equation of motion (\ref{2.7}) as
\begin{equation}
G^{linear}_{\mu \nu} + C^{linear}_{\mu \nu} = -8 \pi \thinspace G \, \left(T_{\mu \nu} + \frac{1}{8 \pi G} (G^{non-linear}_{\mu \nu} +
C^{non-linear}_{\mu \nu}) \right).
\end{equation}
We have already remarked that the left side is divergenceless. Thus we can identify a symmetric and conserved energy-
momentum (pseudo) tensor as
\begin{equation}
\tau_{\mu \nu} = T_{\mu \nu} + \frac{1}{8\pi G} (G^{non-linear}_{\mu \nu} + C^{non-linear}_{\mu \nu}).
\label{2.26}
\end{equation}
It is striking that this structure is present in a theory that seems to violate Lorentz invariance!

In Ref. \cite{rj8} there is a survey of other gravitational energy-momentum (pseudo) tensors for Einstein's theory that
differ from each other by super potentials. In particular there is described a Noether construction with a Belinfante
improvement, which also yields a symmetric, conserved energy-momentum (pseudo) tensor tied to the Poincar\'{e} invariance of the Einstein
theory. It would be interesting to reconsider this construction in the extended theory and to compare the result to (\ref{2.26}).

\section*{Conclusion}
Measuring the intensity of polarized gravity waves is not feasible. Thus for present days, our model is only a theoretical
exercise. Nevertheless, it shows interesting and unexpected behavior in that an important symmetry--diffeomorphism
invariance--is not present in the action, but is restored by the equations of motion. Correspondingly the physical effects
of the symmetry breaking are quite hidden.

An analogy can be made with the St\"{u}ckelberg formalism for massive, Abelian gauge fields. The action
\begin{equation}
I_m = \int d^4 x \left(-\frac{1}{4} F^{\mu \nu} F_{\mu \nu} + \frac{1}{2} m^2 A^\mu A_\mu
\right) \nonumber
\end{equation}
is not gauge invariant
\begin{equation}
\delta_{\mbox{\scriptsize gauge}} \ I_m = \int \ m^2 A^\mu \partial_\mu \lambda = -\int  m^2 \partial_\mu A^\mu \lambda, \nonumber
\end{equation}
but is seen to be broken by $\partial_\mu A^\mu$. However, the equation of motion
\begin{equation}
\partial_\mu F^{\mu \nu} + m^2  A^\nu = J^\nu \nonumber
\end{equation}
has as a consequence (for conserved matter currents) $\partial_\nu A^\nu = 0$, 
and the equation may be presented in gauge invariant form.
\begin{equation}
\partial_\mu F^{\mu \nu} + m^2 \left(g^{\mu \nu} - \frac{\partial^\mu \partial^\nu}{\Box}\right) A_\mu = J^\nu \nonumber
\end{equation}

But an important difference remains. If the the mass is promoted to a field, $m^2 \to m^2 (x)$,
and this field is varied, then the resulting equation
$A^\mu A_\mu = 0$,
is not consequent to the original equation of motion. Indeed it is an unacceptable equation, because
it prevents finding non trivial solutions. (One recognizes the Higgs mechanism, in unitary gauge,
as providing a kinetic term and a potential for the field $``m^2 (x)"$, so that its equation 
of motion becomes dynamically acceptable.)

\end{document}